# Gate-Tunable Optical Extinction of Graphene Nanoribbon Nanoclusters

*Erin Sheridan[1,2], Gang Li[3], Mamun Sarker[3], Shan Hao[1,2], Ki-Tae Eom[4], Chang-Beom Eom[4], Alexander Sinitskii[3], Patrick Irvin[1,2], Jeremy Levy[1,2]*

[1] Department of Physics and Astronomy, University of Pittsburgh, Pittsburgh, PA 15260, USA

[2] Pittsburgh Quantum Institute, Pittsburgh, PA, 15260, USA

[3] Department of Chemistry, University of Nebraska-Lincoln, Lincoln, NE 68588, USA

[4] Department of Materials Science and Engineering, University of Wisconsin-Madison, Madison, WI 53076, USA

**Abstract** We investigate the optical response of graphene nanoribbons (GNRs) using the broadband nonlinear generation and detection capabilities of nanoscale junctions created at the $LaAlO_3/SrTiO_3$ interface. GNR nanoclusters measured to be as small as 1-2 GNRs in size are deposited on the $LaAlO_3$ surface with an atomic force microscope tip. Time-resolved nonlinear optical probes of GNR nanoclusters reveal a strong, gate-tunable second and third harmonic response, as well as strong extinction of visible to near-infrared (VIS-NIR) light at distinct wavelengths, similar to previous reports with graphene.

**Introduction**

Graphene nanoribbons (GNRs), quasi-one dimensional honeycomb arrangements of carbon with precisely defined chemical makeup defined by synthetic chemistry, have emerged as a system of interest in low-dimensional condensed matter physics. On account of their high electronic mobility[1], high thermal conductivity[2], and low noise[3], they are also candidates for use in next-generation integrated circuits and other systems[4-6]. This is in part due to their unique electronic structure. Unlike pristine two-dimensional graphene sheets, GNRs often have energy band gaps[7,8]; furthermore, their electronic structure depends sensitively on their width, edge geometry, and dopants[9-12]. This allows for atomic-scale engineering of diverse physical properties[13, 14].

GNRs are known to exhibit unusual optical properties, such as edge-dependent optical selection rules[15], and have been integrated into optoelectronic devices[3, 16]. GNRs also host gate-tunable surface plasmon polaritons (or plasmons)[17-19], which can give rise to a plasmon-enhanced nonlinear optical response[20-23]. The integration of GNRs with other nanostructures opens up the possibility for many new device concepts, such as programmable nanoplasmonic arrays for use in integrated photonic circuits and GNR electron waveguides. These and other devices further advance GNRs as a candidate material in nanophotonic and quantum information applications. Characterizing and integrating a small number of GNRs is necessary, as aggregation effects obscure their intrinsic properties[24], and precise control is required for quantum applications[25].

Integration of single or few GNRs into devices remains a challenge, however, as does characterization of GNRs optical response beyond ensemble measurements[26]. $LaAlO_3/SrTiO_3$ (LAO/STO) nanostructures are able to characterize the nonlinear optical response of materials like graphene[27], and have a variety of interesting optical and electronic properties themselves[28]. A wide range of LAO/STO-based optoelectronic devices have been created using conductive atomic force microscope lithography[29], including 10 nm-scale photodetectors[30] and nanoscale terahertz (THz) sources and detectors with >100 THz bandwidth[31, 32]. Previously, the THz response of LAO/STO nanojunctions has been coupled to the plasmonic degrees of freedom in single gold plasmonic nanorods[33]. Graphene has recently been integrated with LAO/STO nanostructures as well[34-36], and graphene/LAO/STO nanojunctions exhibit gate-tunable, >99.9% extinction of visible-to-near-infrared (VIS-NIR) light and an enhanced nonlinear optical response[27].

In this Letter we perform nonlinear optical spectroscopy of GNR nanoclusters using nanoscale junctions defined at the LAO/STO interface. First, we briefly detail a GNR deposition method with which one can controllably deposit a very small number of GNRs on a desired substrate with nanoscale resolution. We then discuss results obtained by using this method to integrate GNR nanoclusters into LAO/STO nanojunction devices. Specifically, time-domain photovoltage measurements reveal strong, spectrally sharp, gate-tunable extinction features at VIS-NIR frequencies in addition to an enhanced second harmonic generation (SHG) and third harmonic generation (THG) response. In the discussion section we will compare the observed features in the GNR/LAO/STO nanojunctions to those in graphene/LAO/STO nanojunctions and offer an interpretation of our experimental results.

**Experimental Methods**

The GNRs studied in this work are solution-synthesized semiconducting chevron-type GNRs. These have a bandgap of about 2 eV, which is notably smaller than other chevron GNRs. A powder of synthesized GNRs is created by annealing the synthesized solution in ultra-high vacuum. See Liu *et al.* (2020)[37] for growth and characterization details.

Small clusters of GNRs are placed on the surface of a 3.4 unit cell LAO/STO heterostructure using a process that is summarized in **Figure 1 (b-e)** and described in detail below[38-43]. First, some GNR powder is directly placed onto an LAO/STO substrate. A clean AFM tip is lowered into the GNR powder and raster scanned to pick up a small number of GNRs. The GNR-covered AFM tip is subsequently brought into contact with the LAO surface in the desired "canvas" region. To deposit few-GNR clusters, the tip is raster scanned in contact mode, shedding GNRs as it moves back and forth. Contact-mode AFM scanning can also be used to clean away unwanted GNR clusters. The GNR-coated AFM tip is then removed and replaced with a clean AFM tip, which is used to image the LAO surface in AC or tapping mode. In order to identify GNRs, the AFM image is compared with a reference image taken before depositing the GNRs. An AFM image of deposited GNR clusters is shown in **Figure 2(a)**. Analysis of the AFM image, detailed in the **Supplementary Information**, reveals a distribution of GNR clusters as small as 1-2 GNRs. The GNR cluster used for this investigation (shown in **Figure 2(a,b)** is estimated to contain roughly 10 GNRs.

A LAO/STO nanojunction device is created by first locating a GNR cluster using an AFM. A nanojunction device that surrounds the single GNR nanocluster is then written using c-AFM lithography[33]. The nanojunction consists of a conducting LAO/STO nanowire with a nanoscale (~10 nm) insulating gap enables tunable electric fields on the order of 1 MV/cm to be applied directly to the GNRs. The GNR nanocluster is located on the LAO surface directly above the nanojunction gap, as shown in **Figure 1(f,g)** and **Figure 2(b)**. A nearby looped nanowire serves as a "side gate" to adjust the chemical potential in the device[44, 45]. Additional details of c-AFM lithography and LAO/STO sample growth can be found in the **Supplementary Information**.

After the device is written, the sample is transferred to an optical cryostat (Montana Instruments cryostation), where it is pumped to high vacuum and cooled to a temperature between 5-50 K. The optical response of the GNRs is measured by the same LAO/STO nanowire junction, which behaves as a broadband near-field photodetector[30]. As shown in **Figure 2(c)**, a source-drain bias ($V_{SD}$) is applied across the nanostructure, which is illuminated by ultrafast pulses. The induced photovoltage across the LAO/STO nanojunction, $\Delta V(\tau) = V_+(\tau) - V_-(\tau)$, is measured as a function of the time delay $\tau$ between two pulses.

LAO/STO nanojunctions have been shown to locally generate and detect THz emission[46], with greater than 100 THz bandwidth, via the third-order nonlinear optical process in STO. $V_{SD}$ creates a quasi-static electric field $\vec{E}_{SD}$ across the junction that is highly confined in space to ~10 nm, while input optical fields $\vec{E}_{opt}(\omega_1), \vec{E}_{opt}(\omega_2)$ are sharply peaked in the time domain. The three electric fields mix to generate the nonlinear response of the nanojunction[32]. The power spectrum $S(\Omega)$ versus frequency $\Omega$ is calculated by taking a Fourier transform of the photoresponse $\Delta V(\tau)$ with respect to $\tau$.

**Results**

The optical response of GNR/LAO/STO nanostructures is measured as a function of the side gate bias $V_{sg}$, illustrated in **Figure 3(c)**. Five representative time-domain signals $\Delta V(\tau)$, acquired at different $V_{sg}$ values, are plotted in **Figure 3(a-e)**, along with their corresponding power spectra **Figure 3(f-j)**. Two sharp VIS-NIR extinction features are observed in the LNR response at 395 THz when $V_{sg} = -0.1$ V. An additional extinction feature is observed in the SHG response at $V_{sg} = 0.4$ V. These spectrally sharp extinction features are similar to those observed in graphene/LAO/STO nanojunctions[27]. Additionally, the integrated spectral response over four regions of interest (**Figure 3(k-n)**) reveal correlations between the linear and nonlinear responses of the device. The integrated amplitude of the DFG (0-100 THz) response is correlated with the LNR (300-450 THz) response, while the SHG (675-850 THz) response is correlated with the THG (1050-1250 THz) response. In particular, the SHG and THG are maximal near the $V_{sg}$ value where the VIS-NIR extinction feature appears. Finally, the sign of the envelope of the time-domain photovoltage signal at $V_{sg} = -1.2$ V (**Figure 3(a)**) is opposite to what is measured for $V_{sg} = 1.0$ V (**Figure 3(e)**). A similar sign reversal is also observed as a function of the source-drain bias $V_{SD}$. See the **Supplementary Information** for data from $V_{SD}$-dependent measurements.

To study the power dependence of the extinction feature, an ultrafast pulse shaper was used to vary the input power to the device between 3 µW and 6 µW, while otherwise preserving the temporal profile of the optical pulse. This experiment, summarized in **Figure 4**, shows that the extinction depth and local minimum location exhibits a sensitivity to the power that is similar to what was reported for a graphene/LAO/STO nanojunction, in that the input power can be used to finely tune the extinction ratio and frequency.

A particularly striking example of a VIS-NIR extinction feature in a GNR/LAO/STO device is shown in **Figure 5(a)**. We can estimate the extinction percentage by taking the amplitude of the power spectrum at the extinction frequency (366 THz) and comparing it to the amplitude of the same spectrum at 355 THz, just to the left of the extinction feature. A simple division of the two amplitudes results in a 99.93% estimated extinction of light. **Figure 5(b)** shows a qualitatively similar extinction feature in a graphene/LAO/STO nanojunction device under similar experimental conditions.

**Discussion**

Graphene integrated with LAO/STO nanojunctions exhibits gate-tunable, >99.9% extinction of VIS-NIR light and an associated enhanced nonlinear optical response[27]. Similar VIS-NIR extinction features appear in GNR/LAO/STO nanojunctions (**Figure 5(a)**) and graphene/LAO/STO nanojunctions (**Figure 5(b)**), under similar experimental conditions, and exhibit similar gate-dependent and power-dependent behavior. What's more, as shown in **Figure 3**, the SHG and THG responses are both maximized when the VIS-NIR extinction feature appears. It is important to note that both the SHG and THG responses are enhanced, despite the fact that SHG is an even-ordered harmonic and THG is an odd-ordered harmonic. This is in contrast to previous work, in which only THG is enhanced[23].

The exact physical mechanism underlying the remarkable optical extinction in both graphene and GNR/LAO/STO nanojunctions is still unresolved. However, the fact that such similar behavior is observed in both systems indicates that the mechanism in graphene is localized at the junction and very likely involves the generation of gate-tunable plasmons. LAO/STO nanojunctions confine a ~1 V source-drain bias to a ~ 10 nm region, creating extremely large dipole electric fields directly beneath the GNR nanocluster. Such strong electric fields, on the order of 1 MV/cm, should strongly couple to the plasmonic modes of the GNRs and push them to the highly gated regime. It has been shown that in this regime, nanostructured graphene can host VIS-NIR plasmons[47], and in that case, plasmon absorption should lead to extinction of VIS-NIR light at the plasmon energy. The clear resemblance between the graphene and GNR LAO/STO nanojunction responses implies that the extinction features in graphene are likely the result of a confined GNR-like structure induced by the large electric field.

Graphene plasmons concentrate light into nanometer-scale volumes, significantly intensifying the electric fields upon which nonlinear optical phenomena depend[21, 22, 48, 49]. Previous theoretical results have shown that doped graphene nanostructures exhibit plasmon-assisted high harmonic generation at odd and even harmonics[20]. Therefore, the observed robust SHG and THG in GNRs could be attributed to a plasmon-enhanced nonlinear optical response. It is important to note that in graphene/LAO/STO nanojunctions, the DFG response is maximized when VIS-NIR extinctions appear[27], while in GNR/LAO/STO nanojunctions, the DFG enhancement occurs elsewhere. Planar graphene has been shown to exhibit plasmon-enhanced optical rectification when inversion symmetry is broken [50]. This effect is not necessarily present in GNRs, which could explain why the behavior of the DFG response varied between graphene and GNR nanocluster junctions.

Finally, as shown in **Figure 3**, the sign of the envelope of the time domain signal changes at a particular $V_{sg}$ value. Switching behavior is also observed as a function of the source-drain bias $V_{SD}$. One possible explanation could be that the sign change is associated with a doping change within the GNRs from *n*-type to *p*-type[51-54].

Although GNRs may play an important role in next-generation electronics, photonics, and possibly quantum information applications[25, 55], integrating single- or few-GNRs into nanoscale devices remains difficult[11, 56], as is achieving desired electrical performance[52]. The deposition technique utilized in this work allows for the placement and integration of GNR clusters as small as 1-2 GNRs in size with nanoscale control. Furthermore, c-AFM-defined LAO/STO nanostructures can effectively contact and gate the deposited clusters without requiring complex nanofabrication methods, and gate-dependent optical studies reveal VIS-NIR extinction features and strong optical nonlinearities. Indeed, >99.9% extinction of light is achieved in a GNR nanocluster in a frequency range where graphene typically absorbs only $\approx$ 2% of light[57].

**Conclusions**

We have performed nonlinear optical spectroscopy on GNR nanoclusters deposited on LAO/STO. GNR/LAO/STO nanojunctions are shown to exhibit gate-tunable, narrow-band, near-total (> 99.9 %) extinction of light across a broad range of VIS-NIR frequencies as well as strong second-order and third-order optical nonlinearities. The observed extinction features and nonlinear response bear a strong resemblance to those observed in graphene/LAO/STO junctions. The integration of GNRs with LAO/STO nanostructures opens up the possibility for many new device concepts, such as programmable nanoplasmonic arrays and GNR/LAO/STO electron waveguides. These and other devices further advance GNRs as a candidate material in nanophotonic and quantum information applications.

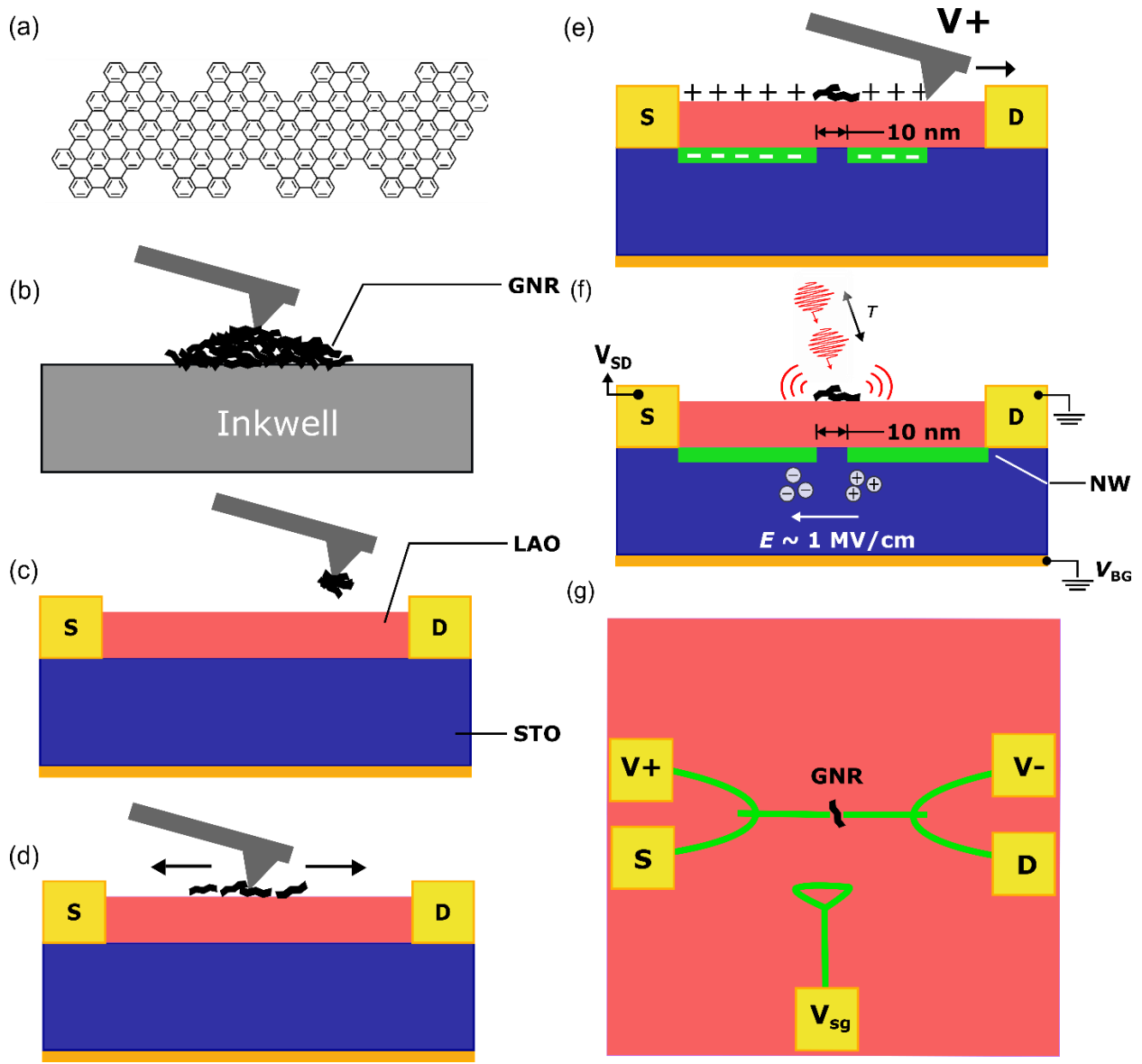

**Figure 1. GNR nanocluster deposition and c-AFM lithography**. (a) Diagram of the chevron a-GNRs used in this work. (b) GNR "inkwell" consisting of a GNR powder on an LAO/STO substrate. An AFM tip is brought into contact with the inkwell and comes away with a small number of GNRs. (c) The tip with a small number of GNRs is brought to a clean LAO/STO sample. (d) The tip is scanned along the LAO surface in contact mode and sheds GNR nanoclusters. (e) A clean AFM tip is used to create a GNR/LAO/STO nanojunction via c-AFM lithography. (f) Time-domain optical measurements are performed on the gated GNR

nanojunction device. +, - charges represent the dipole field established across the junction. (g) Top-view diagram of four-terminal nanojunction and nearby side gate nanowire.

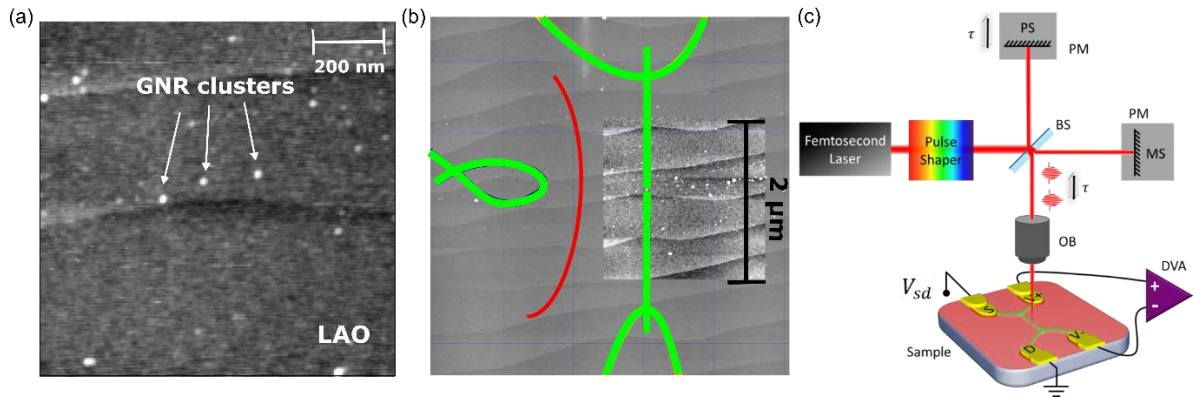

**Figure 2. GNR/LAO/STO nanojunction device.** (a) AFM image of GNR clusters deposited on LAO surface. The rightmost labeled cluster is used for the device in panel (b). (b) AFM image of GNR/LAO/STO nanojunction. A GNR nanocluster is located in the center of the nanojunction gap on the LAO surface. The red line represents a negative-voltage wire which ensures that there is no leakage between side gate and nanojunction. (c) Diagram of optical measurement setup. BS: beam splitter, PM: plane mirror, MS: mechanical stage, PS: piezoelectric stage, OB: objective, and DVA: differential voltage amplifier. The dimensions are not to scale.

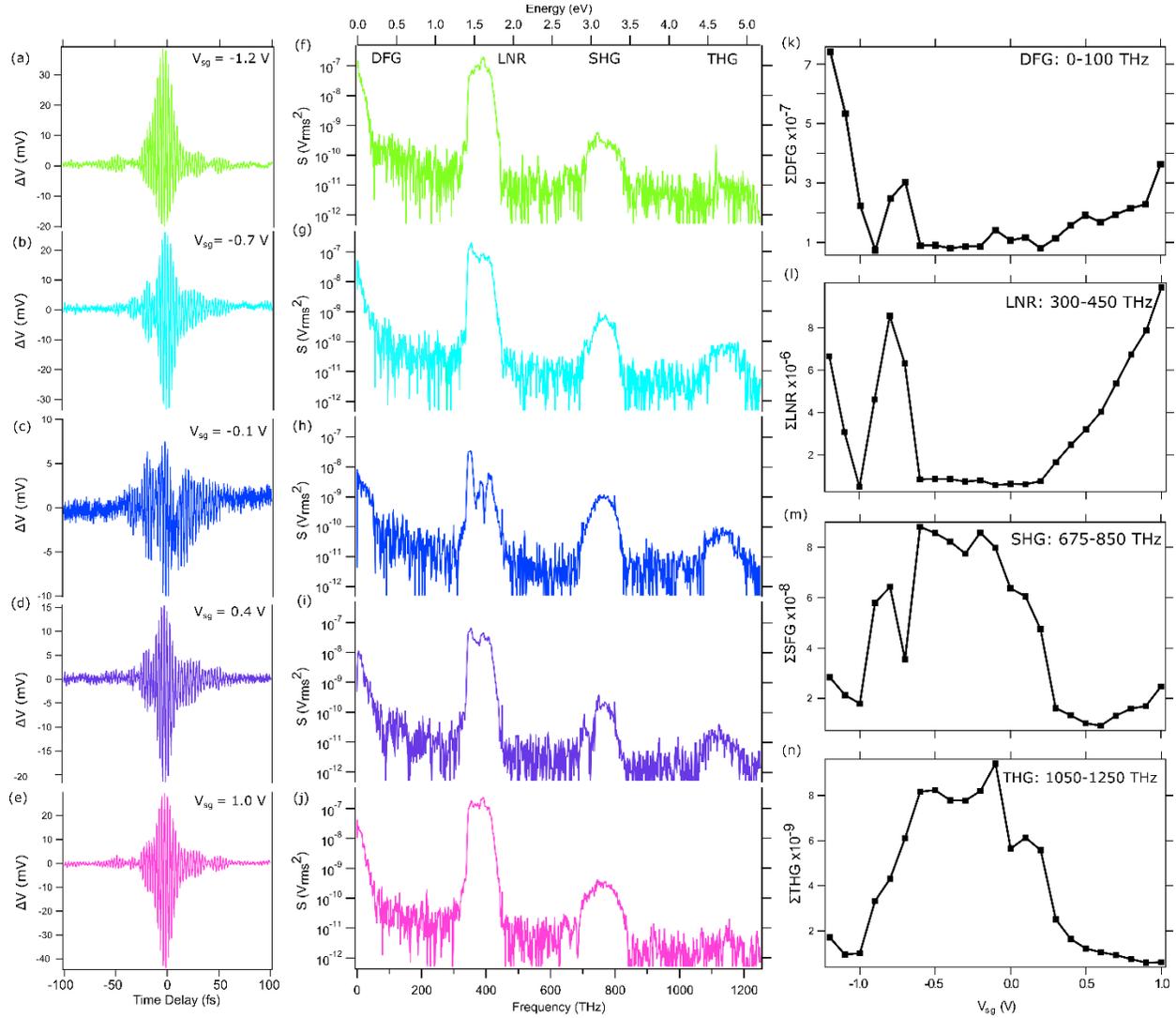

**Figure 3. Side-gate dependence of GNR nanojunction response.** (a-e) Time-domain signals and (f-j) corresponding power spectra at 5 different side gate values. Power spectra regions are labeled in panel (f). Integrals of the (k) DFG, (l) LNR, (m) SHG and (n) THG responses reveal correspondence between the DFG and LNR response, and between the SHG and THG response. $T = 5$ K, $V_{SD} = -0.75$ V, input power 8 μW.

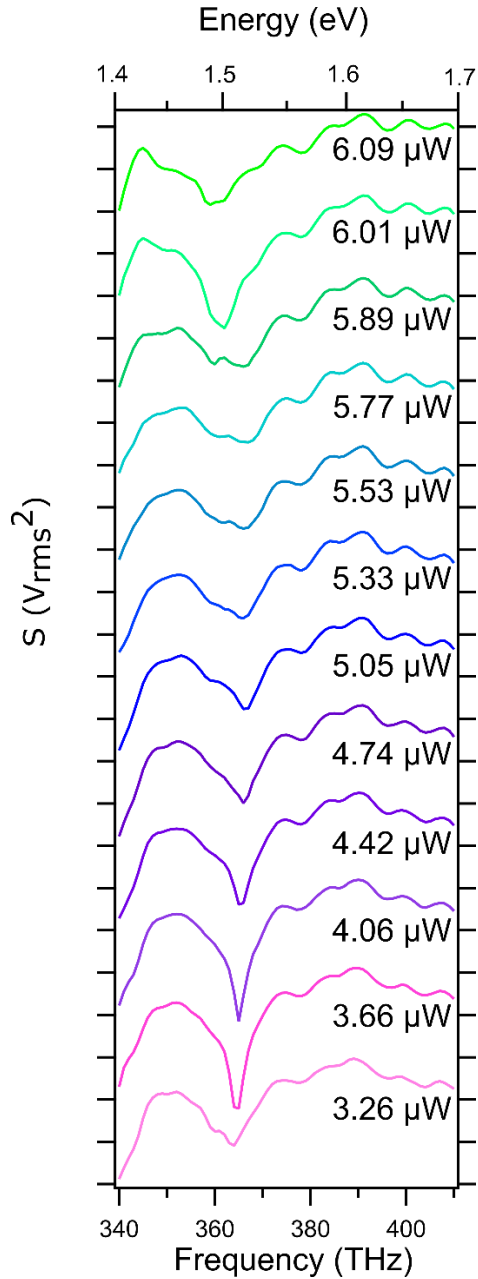

**Figure 4. Input power dependence of VIS-NIR extinction in GNR device**. **(a)** Waterfall plot of VIS-NIR region of power spectra zoomed in to the vicinity of an extinction feature. Each tick mark denotes two order of magnitude and plots are vertically offset for clarity. $T = 50$ K, $V_{SD} = -500$ mV, $V_{sg} = 0$ V.

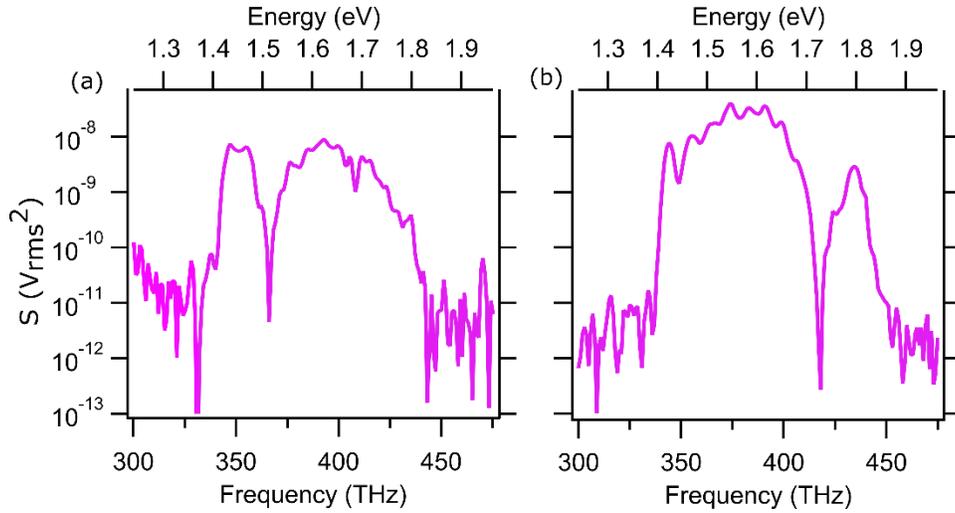

**Figure 5. VIS-NIR extinction in GNR nanoclusters and graphene**. VIS-NIR extinction features in (a) GNR/LAO/STO nanojunction ($T = 5$ K, $V_{SD} = -25$ mV, $V_{sg} = -350$ mV) and (b) graphene/LAO/STO nanojunction ($T = 10$ K, $V_{SD} = -1.25$ V).

**Author Contributions**

E.S. performed experiments and analyzed data. G.L. and M.S. grew the GNRs. K.E. grew the LAO/STO samples. S.H. processed and patterned the LAO/STO samples. C.B.E., A.S., P.I. and J.L. directed and supervised the project. All authors contributed to the writing of the manuscript. All authors have given approval to the final version of the manuscript.

**Funding acknowledgements**

J.L. acknowledges support from the Office of Naval Research, grant N00014-20-1-2481.  E.S. acknowledges support from the National Science Foundation GRFP under Grant No. 1747452. The synthesis of GNRs was supported by the Office of Naval Research (N00014-19-1-2596) and the National Science Foundation through CHE-1455330.The research at UW-Madison is funded by the Gordon and Betty Moore Foundation's EPiQS Initiative, Grant GBMF9065 to C.B.E.

**Data availability**

The data that support the findings of this study are available from the corresponding author upon reasonable request.